\documentclass[11pt,twoside]{article}

\usepackage{asp2010}

\resetcounters

\begin{document}

\title{Reconnection in turbulent astrophysical fluids}
\author{A. Lazarian,$^1$ G. Eyink,$^2$ E. Vishniac,$^3$ and G. Kowal$^4$}
\affil{$^1$Department of Astronomy, University of Wisconsin, Madison, WI 53706, US}
\affil{$^2$Department of Applied Mathematics and Statistics, The Johns Hopkins University, Baltimore, Maryland 21218, US},
\affil{$^3$Department of Physics and Astronomy, McMaster University, 1280 Main Street West, Hamilton, Ontario L8S 4M1, Canada},
\affil{$^4$Escola de Artes, Ci\^{e}ncias e Humanidades, Universidade de S\~{a}o Paulo, Av. Arlindo B\'{e}ttio, 1000 -- Ermelino Matarazzo, CEP 03828-000, S\~{a}o Paulo, SP, Brazil}

\begin{abstract}
Magnetic reconnection is a fundamental process of magnetic field topology
change. We analyze the connection of this process with turbulence which is
ubiquitous in astrophysical environments. We show how Lazarian \& Vishniac
(1999) model of turbulent reconnection is connected to the experimentally proven
concept of Richardson diffusion and discuss how turbulence violates the
generally accepted notion of magnetic flux freezing. We note that in
environments that are laminar initially turbulence can develop as a result of
magnetic reconnection and this can result in flares of magnetic reconnection in
magnetically dominated media. In particular, magnetic reconnection can initially
develop through tearing, but the transition to the turbulent state is expected
for astrophysical systems.
\end{abstract}

\section{Introduction}

Magnetic fields modify fluid dynamics and it is generally believed that magnetic
fields embedded in a highly conductive fluid retain their topology for all time
due to the magnetic fields being frozen-in \citep{Alfven:1943, Parker:1979}.
Nevertheless, highly conducting ionized astrophysical objects, like stars and
galactic disks, show evidence of changes in topology, i.e. ``magnetic
reconnection'', on dynamical time scales \citep{Parker:1970, Lovelace:1976,
PriestForbes:2002}.

In this short review we argue that the concept of magnetic flux being frozen
must be seriously modified to account for turbulence, which induces fast
magnetic reconnection.  The focus of the discussion in this review is
\cite[][henceforth LV99]{LazarianVishniac:1999} model of reconnection in
turbulent fluids\footnote{We discuss weakly turbulent fluids which is a natural
generalization of a traditional Sweet-Parker treatment of laminar fluxes.  Thus
LV99 should not be understood as the scheme of reconnection of magnetic field
which is violently bend at all scales up to the dissipation one.} and its
consequences.  A more detailed discussion of this phenomenon can be found in a
our extensive review, i.e. in \cite{LazarianEyinkVishniacKowal:2014}.  We may
add that some aspects of turbulent reconnection are illuminated in other recent
reviews i.e. \cite{BrowningLazarian:2013} and \cite{KarimabadiLazarian:2013}.
The first one analyzes the reconnection in relation to solar flares, the other
provides the comparison of the PIC simulations of the reconnection in
collisionless plasmas with the reconnection in turbulent MHD regime. At the same
time we would like to stress that while in \cite{KarimabadiLazarian:2013} the
solar wind data analysis is presented as supporting the importance of plasma
effects in reconnection, our present work in progress shows, on the contrary,
good correspondence of the solar wind reconnection with the predictions of LV99
model.

\section{Turbulent reconnection: history of ideas}

The first attempts to appeal to turbulence in order to enhance the reconnection
rate were made more than 40 years ago.  For instance, some papers have
concentrated on the effects that turbulence induces on the microphysical level
\citep{Speiser:1970, JacobsonMoses:1984}.  However, these effects are
insufficient to produce reconnection speeds comparable to the Alfv\'{e}n speed
in most astrophysical environments.

``Hyper-resistivity'' \citep{Strauss:1986, BhattacharjeeHameiri:1986,
HameiriBhattacharjee:1987, DiamondMalkov:2003} is a more subtle attempt to
derive fast reconnection from turbulence within the context of mean-field
resistive MHD. The form of the parallel electric field can be derived from
magnetic helicity conservation.  Nevertheless, integrating by parts one obtains
a term which looks like an effective resistivity proportional to the magnetic
helicity current.  In addition, there are several assumptions implicit in this
derivation.  The most important objection to this approach is that by adopting a
mean-field approximation, one is already assuming some sort of small-scale
smearing effect, equivalent to fast reconnection.  Furthermore, the integration
by parts involves assuming a large scale magnetic helicity flux through the
boundaries of the exact form required to drive fast reconnection. The problems
of the hyper-resistivity approach are discussed in detail in LV99 as well as in
\cite{EyinkLazarianVishniac:2011}. We believe that this this approach is not
fruitful.

A more productive development was related to studies of instabilities of the
reconnection layer.  \cite{Strauss:1988} examined the enhancement of
reconnection through the effect of tearing mode instabilities within current
sheets.  However, the resulting reconnection speed enhancement is roughly what
one would expect based simply on the broadening of the current sheets due to
internal mixing.  We note, however, that in a more recent work
\cite{ShibataTanuma:2001} extended the concept suggesting that tearing may
result in fractal reconnection taking place on very small scales.
\cite{Waelbroeck:1989} considered not the tearing mode, but the resistive kink
mode to accelerate reconnection.  The numerical studies of tearing have become
an important avenue for more recent reconnection research
\citep{LoureiroUzdenskySchekochihinCowleyYousef:2009,
BhattacharjeeHuangYangRogers:2009}.  As we discuss later in realistic 3D
settings tearing instability develops turbulence \citep{KarimabadiLazarian:2013,
Beresnyak:2013}) which induces a transfer from laminar to turbulent
reconnection\footnote{Also earlier works suggest such a transfer
\citep{DahlburgAntiochosZang:1992, DahlburgKarpen:1994, Dahlburg:1997,
FerraroRogers:2004}}.

Finally, a study of 2D magnetic reconnection in the presence of external
turbulence was done by \cite{MatthaeusLamkin:1985, MatthaeusLamkin:1986}.  An
enhancement of the reconnection rate was reported, but the numerical setup
precluded the calculation of a long term average reconnection rate.  One may
argue that bringing in the Sweet-Parker model of reconnection magnetic field
lines closer to each other one can enhance the instantaneous reconnection rate,
but this does not mean that averaged long term reconnection rate increases.
This, combined with the absence of the theoretical predictions of the expected
reconnection rates makes it difficult to make definitive conclusions from the
study.  Note that, as we discussed in \cite{EyinkLazarianVishniac:2011}, the
nature of turbulence is different in 2D and 3D. Therefore, the effects
accelerating magnetic reconnection mentioned in the aforementioned 2D studies,
i.e. formation of X-points, compressions, may be relevant for 2D set ups, but
not relevant for the 3D astrophysical reconnection.  We stress, that these
effects are not invoked in the model of the turbulent reconnection that we
discuss below.

In a sense, the above study is the closest predecessor of LV99 work that we deal
below.  However, there are very substantial differences between the approach of
LV99 and \cite{MatthaeusLamkin:1985}.  For instance, LV99, as is clear from the
text below, uses an analytical approach and, unlike \cite{MatthaeusLamkin:1985},
(a) provides analytical expressions for the reconnection rates; (b) identifies
the broadening arising from magnetic field wandering as the mechanism for
inducing fast reconnection; (c) deals with 3D turbulence and identifies
incompressible Alfv\'{e}nic motions as the driver of fast reconnection.

\section{3D reconnection in weakly turbulent fluid}

\begin{figure}
\centering
\includegraphics[width=0.5\textwidth]{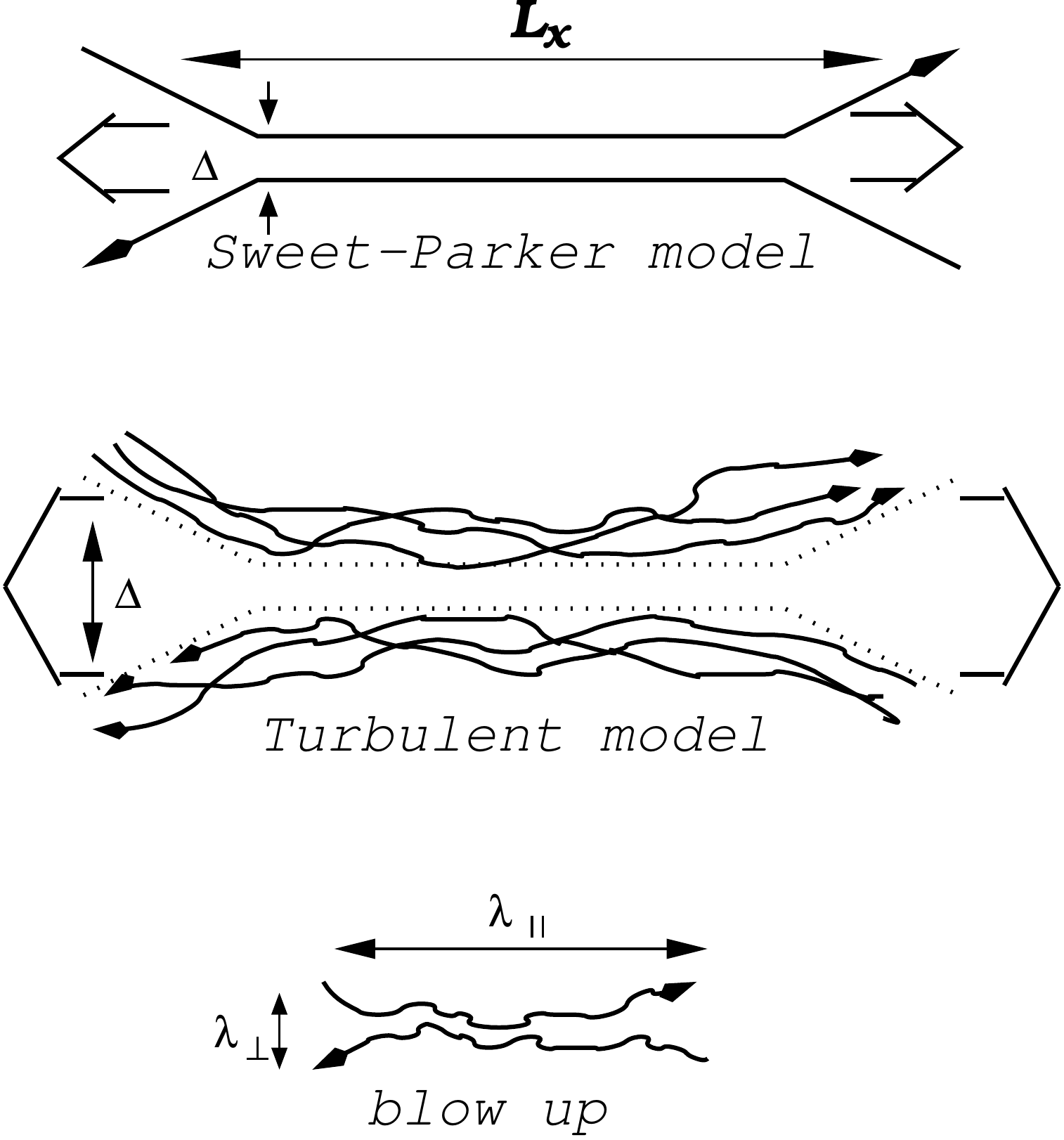}
\caption{{\it Upper plot}: Sweet-Parker model of reconnection.  The outflow is
limited to a thin width $\delta$, which is determined by Ohmic diffusivity.  The
other scale is an astrophysical scale $L \gg \delta$.  Magnetic field lines are
assumed to be laminar.
{\it Middle plot}: Turbulent reconnection model that accounts for the
stochasticity of magnetic field lines.  The stochasticity introduced by
turbulence is weak and the direction of the mean field is clearly defined.  The
outflow is limited by the diffusion of magnetic field lines, which depends on
macroscopic field line wandering rather than on microscales determined by
resistivity.
{\it Low plot}: An individual small scale reconnection region.  The reconnection
over small patches of magnetic field determines the local reconnection rate. The
global reconnection rate is substantially larger as many independent patches
reconnect simultaneously.  Conservatively, the LV99 model assumes that the small
scale events happen at a slow Sweet-Parker rate.  Following
\cite{LazarianVishniacCho:2004}.}
\label{fig1:lv99}
\end{figure}

As astrophysical turbulence is ubiquitous, considering astrophysical reconnection in laminar
environments is not realistic.  As a natural generalization of the
Sweet-Parker model it is appropriate to consider 3D magnetic field wandering
induced by turbulence as in LV99.  The corresponding model of magnetic
reconnection is illustrated by Figure~\ref{fig1:lv99}.

Like the Sweet-Parker model, the LV99 model deals with a generic configuration,
which should arise naturally as magnetic flux tubes try to make their way one
through another.  This avoids the problems related to the preservation of wide
outflow which plagues attempts to explain magnetic reconnection via
Petscheck-type solutions \citep{Petschek:1964}.  In this model if the outflow of
reconnected flux and entrained matter is temporarily slowed down, reconnection
will also slow down, but, unlike Petscheck solution, will not change the nature
of the solution.

The major difference between the Sweet-Parker model and the LV99 model is that
while in the former the outflow is limited by microphysical Ohmic diffusivity,
in the latter model the large-scale magnetic field wandering determines the
thickness of outflow.  Thus LV99 model does not depend on resistivity and,
depending on the level of turbulence, can provide both fast and slow
reconnection rates.  This is a very important property for explaining
observational data related to reconnection flares.

Ultimately, the magnetic field lines will dissipate due to microphysical
effects, e.g. Ohmic resistivity.  However, it is important to understand that in
the LV99 model only a small fraction of any magnetic field line is subject to
direct Ohmic annihilation.  The fraction of magnetic energy that goes directly
into heating the fluid approaches zero as the fluid resistivity vanishes.

The rates obtained in LV99 study are:
\begin{equation}
V_{rec}\approx V_A\min\left[\left({L_x\over L_i}\right)^{1/2},
\left({L_i\over L_x}\right)^{1/2}\right]
{\cal M}_A^2,
\label{eq:lim2a}
\end{equation}
where $V_A {\cal M}_A^2$ is proportional to the turbulent eddy speed.  This
limit on the reconnection speed is fast, both in the sense that it does not
depend on the resistivity, and in the sense that it represents a large fraction
of the Alfv\'{e}n speed when $L_i$ and $L_x$ are not too different and ${\cal
M}_A$ is not too small.  At the same time, Eq. (\ref{eq:lim2a}) can lead to
rather slow reconnection velocities for extreme geometries or small turbulent
velocities.

\section{LV99 model and Richardson diffusion}
\label{sec:richardson}

Magnetic reconnection in LV99 model was described using magnetic field wandering
which, in general, also represents the snapshot of the Richardson diffusion of
magnetic field in space \cite[see][]{EyinkLazarianVishniac:2011}.  The
phenomenon of Richardson diffusion is easy to understand.  At the scales less
that the scale of injection of the strong MHD turbulence, i.e. for $l<l_{trans}$
\cite[LV99, ][]{Lazarian:2006},
\begin{equation}
l_{trans}\sim L(V_L/V_A)^2 \equiv L {\cal M}_A^2.
\label{trans}
\end{equation}
 the magnetic field lines exhibit accelerated diffusion.  Indeed, the separation
between two particles $dl(t)/dt\sim v(l)$ for Kolmogorov turbulence is $\sim
\alpha_t l^{1/3}$, where $\alpha$ is proportional to a cube-root of the energy
cascading rate, i.e. $\alpha_t\approx V_L^3/L$ for turbulence injected with
superAlvenic velocity $V_L$ at the scale $L$.  The solution of this equation is
\begin{equation}
l(t)=[l_0^{2/3}+\alpha_t (t-t_0)]^{3/2},
\label{sol}
\end{equation}
which provides Richardson diffusion or $l^2\sim t^3$.  The accelerating
character of the process is easy to understand physically.  Indeed, the larger
the separation between the particles, the faster the eddies that carry the
particles apart.  It is clear that Richardson diffusion is not an ordinary
diffusion with square separation increasing with time, but a superdiffusion that
cannot be described by a simple diffusion equation.

\cite{EyinkLazarianVishniac:2011} re-derived LV99 expressions for the
reconnection rate from the time dependent Richardson diffusion.  This provides a
way to understand the applicability of LV99 approach to more involved cases,
e.g. for the case of fluid with viscosity much larger than magnetic diffusivity,
i.e. the high Prandtl number fluids.  A partially ionized gas is an example of
such a fluids, which is essential to understand in terms of star formation.

In high Prandtl number media the GS95-type turbulent motions decay at the scale
$l_{\bot, crit}$, which is much larger than the scale of at which Ohmic
dissipation gets important. Thus over a range of scales less than $l_{\bot,
crit}$ magnetic fields preserve their identity and are being affected by the
shear on the scale $l_{\bot, crit}$. This is the regime of turbulence described
in \cite{ChoLazarianVishniac:2002} and \cite{LazarianVishniacCho:2004}.  In view
of the findings in \cite{EyinkLazarianVishniac:2011} to establish when magnetic
reconnection is fast and obeys the LV99 predictions one should establish the
range of scales at which magnetic fields obey Richardson diffusion.  It is easy
to see that the transition to the Richardson diffusion happens when field lines
get separated by the perpendicular scale of the critically damped eddies
$l_{\bot, crit}$.  The separation in the perpendicular direction starts with the
scale $r_{init}$ follows the Lyapunov exponential growth with the distance $l$
measured along the magnetic field lines, i.e. $r_{init} \exp(l/l_{\|, crit})$,
where $l_{\|, crit}$ corresponds to critically damped eddies with $l_{perp,
crit}$.  It seems natural to associate $r_{init}$ with the separation of the
field lines arising from Ohmic resistivity on the scale of the critically damped
eddies
\begin{equation}
r_{init}^2=\eta l_{\|, crit}/V_A,
\label{int}
\end{equation}
where $\eta$ is the Ohmic resistivity coefficient.

In this formulation the problem of magnetic line separation is similar to the
anisotropic analog of the \cite{RechesterRosenbluth:1978} problem
\cite[see][]{NarayanMedvedev:2003, Lazarian:2006} and therefore distance to be
covered along magnetic field lines before the lines separate by the distance
larger than the perpendicular scale of viscously damped eddies is equal to
\begin{equation}
L_{RR}\approx l_{\|, crit} \ln (l_{\bot, crit}/r_{init})
\label{RR}
\end{equation}
Taking into account Eq. (\ref{int}) and that
\begin{equation}
l_{\bot, crit}^2=\nu l_{\|, crit}/V_A,
\end{equation}
where $\nu$ is the viscosity coefficient. Thus Eq. (\ref{RR}) can be rewritten
\begin{equation}
L_{RR}\approx l_{\|, crit}\ln Pt
\label{RR2}
\end{equation}
where $Pt=\nu/\eta$ is the Prandtl number.

If the current sheets are much longer than $L_{RR}$, then magnetic field lines
undergo Richardson diffusion and according to \cite{EyinkLazarianVishniac:2011}
the reconnection follows the laws established in LV99.  In other words, on
scales significantly larger than the viscous damping scale LV99 reconnection is
applicable.  At the same time on scales less than $L_{RR}$ magnetic reconnection
may be slow\footnote{Incidentally, this can explain the formation of density
fluctuations on scales of thousands of Astronomical Units, that are observed in
the ISM.}.  Somewhat more complex arguments were employed in
\cite{LazarianVishniacCho:2004} to prove that the reconnection is fast in the
partially ionized gas.  For our further discussion it is important that LV99
model is applicable both to fully ionized and partially ionized plasmas.

\section{Testing of LV99 predictions and Richardson diffusion}
\label{sec:testing}

\begin{figure}
\centering
\includegraphics[height=.45\textheight]{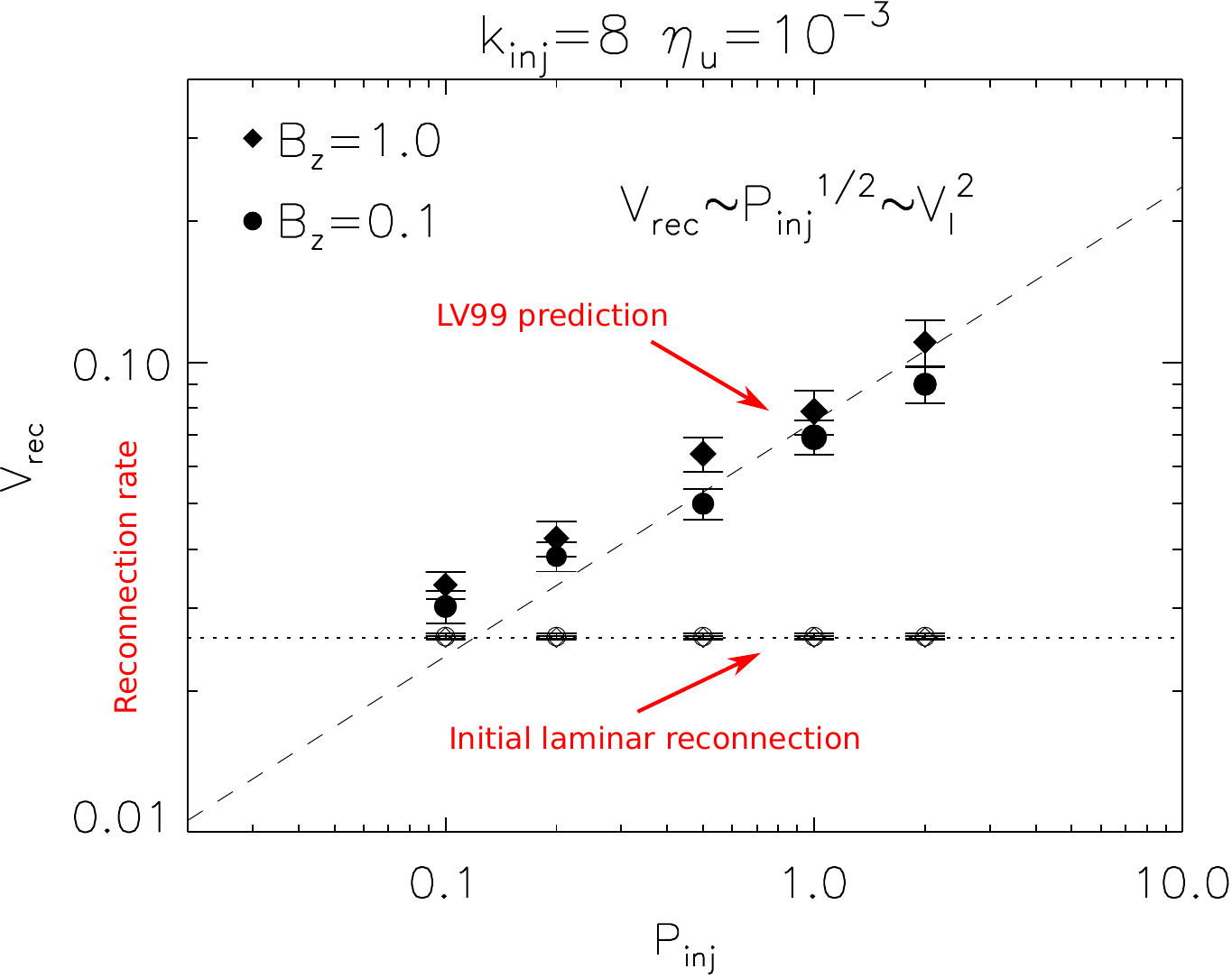}
\caption{The dependence of the reconnection velocity on the injection power for
different simulations with different drivings. The predicted LV99 dependence is
also shown. $P_{inj}$ and $k_{inj}$ are the injection power and scale,
respectively, $B_z$ is the guide field strength, and $\eta_u$ the value of
uniform resistivity coefficient. From
\cite{KowalLazarianVishniacOtmianowska-Mazur:2012}. \label{fig2:lv99dep}}
\end{figure}

Simulations in \cite{KowalLazarianVishniacOtmianowska-Mazur:2009,
KowalLazarianVishniacOtmianowska-Mazur:2012} confirmed no dependence of
turbulent reconnection on resistivity and provided good correspondence to the
LV99 analytical predictions the injection power, i.e. $V_{rec}\sim
P_{inj}^{1/2}$.  The corresponding dependence is shown in
Figure~\ref{fig2:lv99dep}.

As we discussed, the LV99 model is intrinsically related to the concept of
Richardson diffusion in magnetized fluids. Thus by testing the Richardson
diffusion of magnetic field, one also provides tests for the theory of turbulent
reconnection.

\begin{figure}[!ht]
\centering
\raisebox{-0.5\height}{\includegraphics[width=0.49\textwidth]{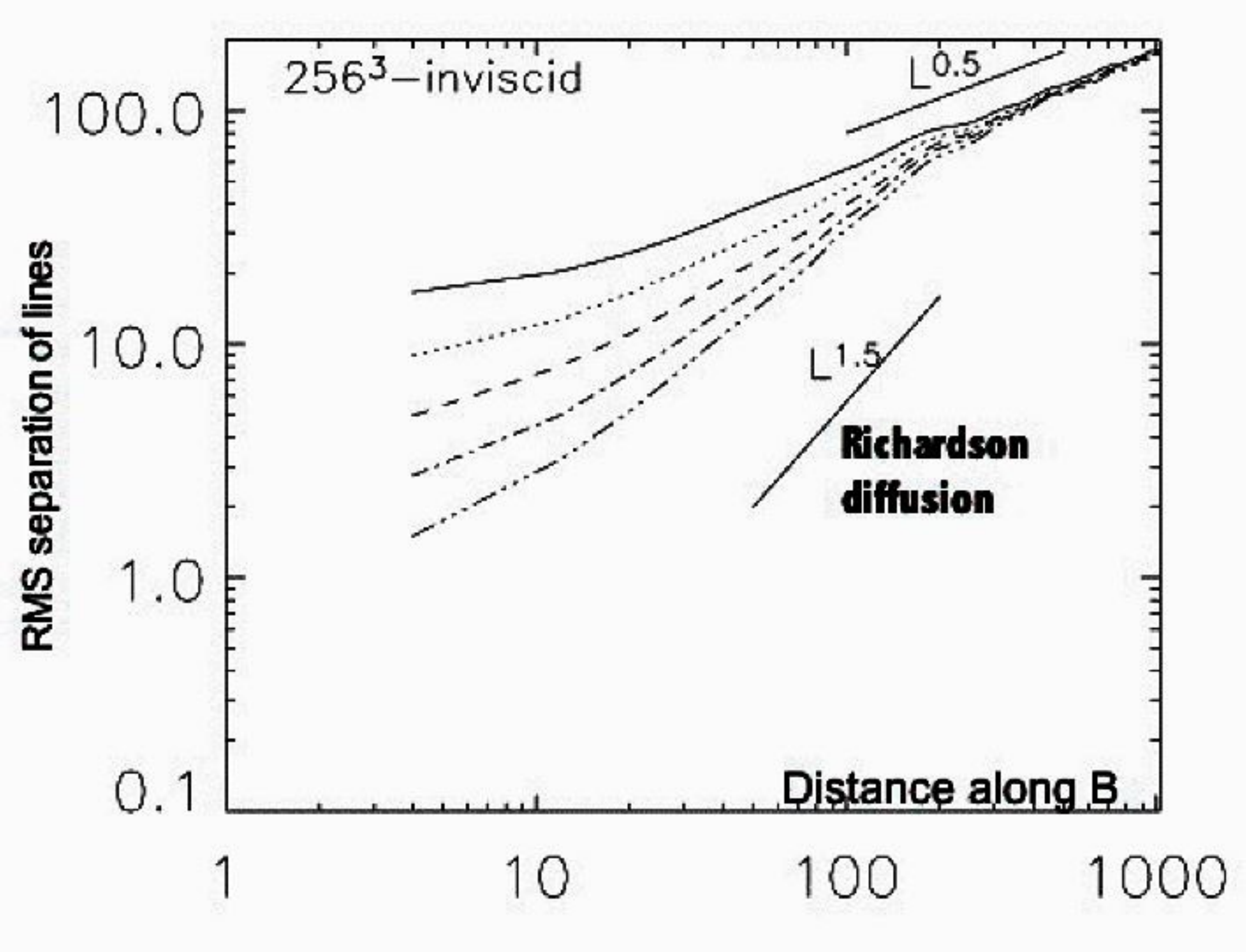}}
\raisebox{-0.5\height}{\includegraphics[width=0.49\textwidth]{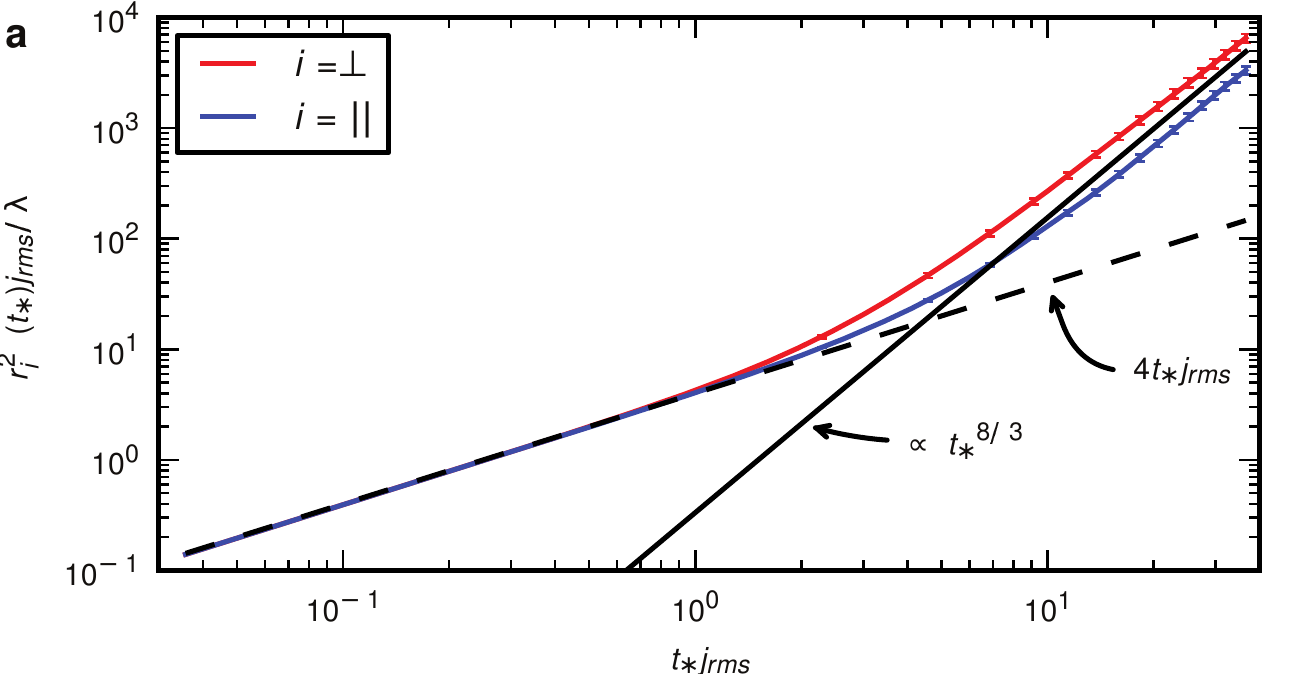}}
\caption{{\it Left panel}. Richardson diffusion of magnetic field lines in
space. From \cite{LazarianVishniacCho:2004} {\it Right panel}. Stochastic
trajectories that arrive at a fixed point in the archived MHD flow, color-coded
red, green, and blue from earlier to later times.  From
\cite{EyinkVishniacLalescuAluieKanovBurgerBurnsMeneveauSzalay:2013}.
\label{fig3:richardson}}
\end{figure}

The first numerical tests of Richardson diffusion were related to magnetic field
wandering predicted in LV99 \cite{MaronChandranBlackman:2004,
LazarianVishniacCho:2004}.  In Figure~\ref{fig3:richardson} we show the results
obtained in \cite{LazarianVishniacCho:2004}.  There we clearly see different
regimes of magnetic field diffusion, including the $y\sim x^{3/2}$ regime.  This
is a manifestation of the spatial Richardson diffusion.

A numerical study of the temporal Richardson diffusion of magnetic field-lines
was performed  in
\cite{EyinkVishniacLalescuAluieKanovBurgerBurnsMeneveauSzalay:2013}.  For this
experiment, stochastic fluid trajectories had to be tracked backward in time
from a fixed point in order to determine which field lines at earlier times
would arrive to that point and be resistively ``glued together''.  Hence, many
time frames of an MHD simulation were stored so that equations for the
trajectories could be integrated backward.  The results of this study are
illustrated in Figure~\ref{fig3:richardson}.  It shows the trajectories of the
arriving magnetic field-lines, which are clearly widely dispersed backward in
time, more resembling a spreading plume of smoke than a single ``frozen-in''
line.

The implication of these results is that standard diffusive motion of
field-lines holds for only a very short time, of order of the resistive time,
and is then replaced by super-diffusive, explosive separation by turbulent
relative advection.  This same effect should occur not only in resistive MHD but
whenever there is a long power-law turbulent inertial range.  Whatever plasma
mechanism of line-slippage holds at scales below the ion gyroradius--- electron
inertia, pressure anisotropy, etc.---will be accelerated and effectively
replaced by the ideal MHD effect of Richardson dispersion.

\section{Observational testing of LV99}

Qualitatively, one can argue that there is observational evidence in favor of
the LV99 model.  For instance, observations of the thick reconnection current
outflow regions observed in the Solar flares \citep{CiaravellaRaymond:2008} were
predicted within LV99 model at the time when the competing plasma Hall term
models were predicting X-point localized reconnection.  However, as plasma
models have evolved since 1999 to include tearing and formation of magnetic
islands \cite[see][]{DrakeOpherSwisdakChamoun:2010} it is necessary to get to a
quantitative level to compare the predictions from the competing theories and
observations.

To be quantitative one should relate the idealized model LV99 turbulence driving
to the turbulence driving within solar flares.  In LV99 the turbulence driving
was assumed isotropic and homogeneous at a distinct length scale $L_{inj}$.  A
general difficulty with observational studies of turbulent reconnection is the
determination of $L_{inj}$.  One possible approach is based on the the relation
$\varepsilon \simeq u_L^4/V_AL_{inj}$ for the weak turbulence energy cascade
rate.  The mean energy dissipation rate $\varepsilon$ is a source of plasma
heating, which can be estimated from observations of electromagnetic radiation
\cite[see more in][]{EyinkLazarianVishniac:2011}.  However, when the energy is
injected from reconnection itself, the cascade is strong and anisotropic from
the very beginning.  If the driving velocities are sub-Alfv\'{e}nic, turbulence
in such a driving is undergoing a transition from weak to strong at the scale $L
{\cal M}_A^2$ (see \S\ref{sec:richardson}).  The scale of the transition
corresponds to the velocity ${\cal M}_A^2 V_A$.  If turbulence is driven by
magnetic reconnection, one can expect substantial changes of the magnetic field
direction corresponding to strong turbulence.  Thus it is natural to identify
the velocities measured during the reconnection events with the strong MHD
turbulence regime.  In other words, one can use:
\begin{equation}
V_{rec}\approx U_{obs, turb} (L_{inj}/L_x)^{1/2},
\label{obs}
\end{equation}
where $U_{obs, turb}$ is the spectroscopically measured turbulent velocity
dispersion. Similarly, the thickness of the reconnection layer should be defined
as
\begin{equation}
\Delta\approx L_x (U_{obs, turb}/V_A) (L_{inj}/L_x)^{1/2}.
\label{delta_obs}
\end{equation}

Naturally, this is just a different way of presenting LV99 expressions, but
taking into account that the driving arises from reconnection and therefore
turbulence is strong from the very beginning (see more in
\cite{EyinkVishniacLalescuAluieKanovBurgerBurnsMeneveauSzalay:2013}. The
expressions given by Eqs.~(\ref{obs}) and (\ref{delta_obs}) can be compared with
observations in \cite{CiaravellaRaymond:2008}.  There, the widths of the
reconnection regions were reported in the range from 0.08$L_x$ up to 0.16$L_x$
while the the observed Doppler velocities in the units of $V_A$ were of the
order of 0.1.  It is easy to see that these values are in a good agreement with
the predictions given by Eq.~(\ref{delta_obs}).

In addition, in \cite{SychNakariakovKarlickyAnfinogentov:2009}, the authors
explaining quasi-periodic pulsations in observed flaring energy releases at an
active region above the sunspot, proposed that the wave packets arising from the
sunspots can trigger such pulsations.  This is exactly what is expected within
the LV99 model. We are not aware of any other model of reconnection that would
predict such a triggering.

The criterion for the application of LV99 theory is that the outflow region is
much larger than the ion Larmor radius $\Delta \gg \rho_i$.  This is definitely
satisfied for the solar atmosphere where the ratio of $\Delta$ to $\rho_i$ can
be larger than $10^6$. Plasma effects can play a role for small scale
reconnection events within the layer, since the dissipation length based on
Spitzer resistivity is $\sim 1$ cm, whereas $\rho_i\sim 10^3$ cm.  However, as
we discussed earlier, this does not change the overall dynamics of turbulent
reconnection.

Reconnection throughout most of the heliosphere appears similar to that in the
Sun.  For example, there are now extensive observations of reconnection jets
(outflows, exhausts) and strong current sheets in the solar wind
\citep{Gosling:2012}.  The most intense current sheets observed in the solar
wind are very often not observed to be associated with strong (Alfv\'enic)
outflows and have widths at most a few tens of the proton inertial length
$\delta_i$ or proton gyroradius $\rho_i$ (whichever is larger).  Small-scale
current sheets of this sort that do exhibit observable reconnection have
exhausts with widths at most a few hundreds of ion inertial lengths and
frequently have small shear angles (strong guide fields)
\citep{GoslingPhanLinSzabo:2007, GoslingSzabo:2008}.  Such small-scale
reconnection in the solar wind requires collisionless physics for its
description, but the observations are exactly what would be expected of
small-scale reconnection in MHD turbulence of a collisionless plasma
\citep{VasquezAbramenkoHaggertySmith:2007}.  Indeed, LV99 predicted that the
small-scale reconnection in MHD turbulence should be similar to large-scale
reconnection, but with nearly parallel magnetic field lines and with
``outflows'' of the same order as the local, shear-Alfv\'enic turbulent eddy
motions.

However, there is also a prevalence of very large-scale reconnection events in
the solar wind, quite often associated with interplanetary coronal mass
ejections and magnetic clouds or occasionally magnetic disconnection events at
the heliospheric current sheet  \citep{PhanGoslingDavis:2009, Gosling:2012}.
These events have reconnection outflows with widths up to nearly $10^5$ of the
ion inertial length and appear to be in a prolonged, quasi-stationary regime
with reconnection lasting for several hours. Such large-scale reconnection is as
predicted by the LV99 theory when very large flux-structures with
oppositely-directed components of magnetic field impinge upon each other in the
turbulent environment of the solar wind. The ``current sheet'' producing such
large-scale reconnection in the LV99 theory contains itself many ion-scale,
intense current sheets embedded in a diffuse turbulent background of weaker (but
still substantial) current. Observational efforts addressed to
proving/disproving the LV99 theory should note that it is this broad zone of
more diffuse current, not the sporadic strong sheets, which is responsible for
large-scale turbulent reconnection. Note that the study
\citep{EyinkVishniacLalescuAluieKanovBurgerBurnsMeneveauSzalay:2013} showed that
standard magnetic flux-freezing is violated at general points in turbulent  MHD,
not just at the most intense, sparsely distributed sheets. Thus, large-scale
reconnection in the solar wind is a very promising area for LV99. The situation
for LV99 generally gets better with increasing distance from the sun, because of
the great increase in scales. For example, reconnecting flux structures in the
inner heliosheath could have sizes up to $\sim$100 AU, much larger than the ion
cyclotron radius $\sim10^3$ km \citep{LazarianOpher:2009}.

The magnetosphere is another example that is under active investigation by the
reconnection community.  The situation there is different, as $\Delta\sim
\rho_i$ is the general rule and we expect plasma effects to be dominant.

\subsection{Self-sustained turbulent reconnection}

If the initial magnetic field configuration is laminar, magnetic reconnection
ought to induce turbulence due to the outflow (LV99).  This effect was confirmed
by observing the development of turbulence both in recent 3D Particle in Cell
(PIC) simulations
\citep{KarimabadiRoytershteynWanMatthaeusDaughtonWuShayLoringBorovskyLeonardisChapmanNakamura:2013}
and 3D MHD simulations \citep{Beresnyak:2013, KowalLazarianFalcetaVishniac:2014}.

\begin{figure}[!ht]
\centering
\includegraphics[width=\textwidth]{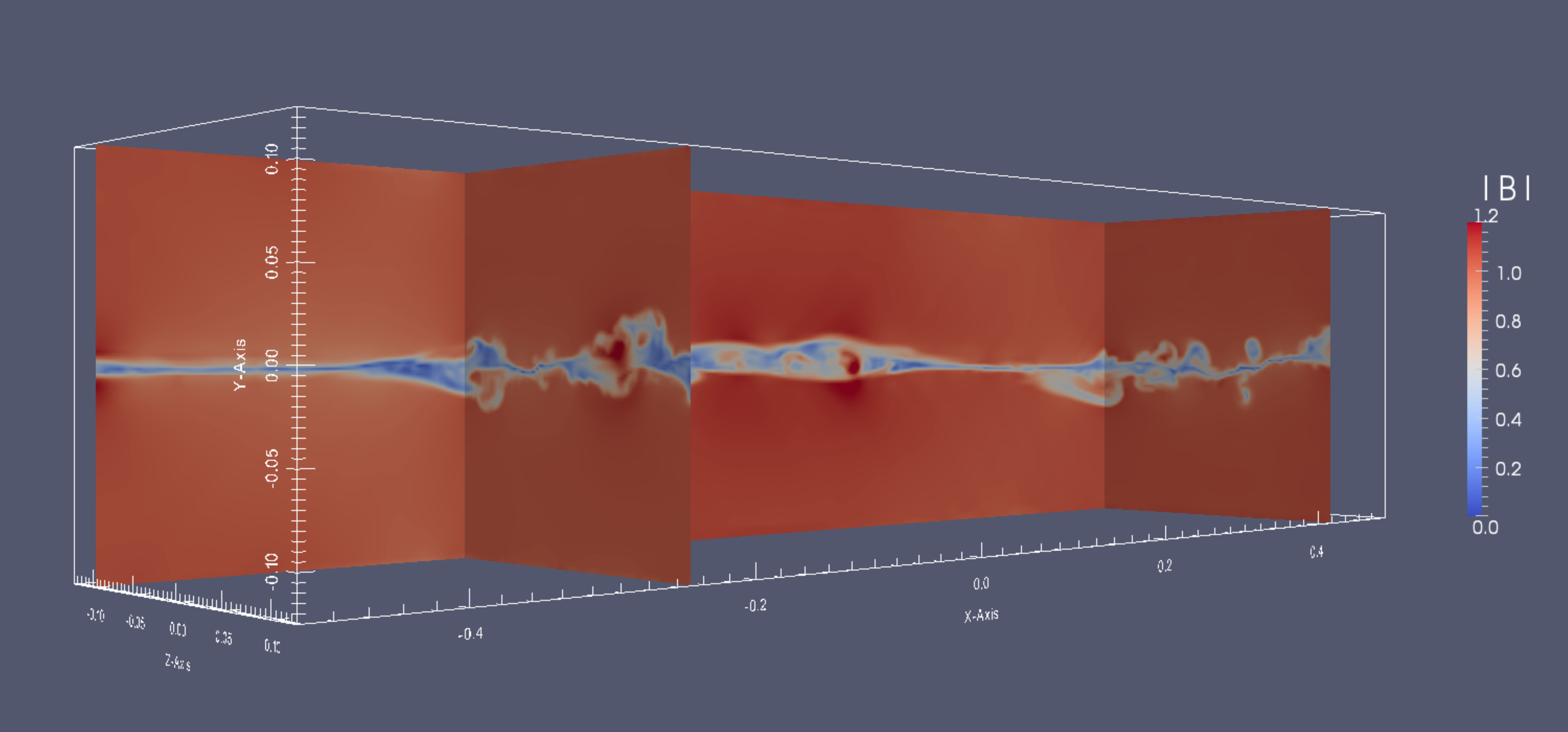}
\caption{Visualization of the model of turbulence generated by the seed
reconnection from \cite{KowalLazarianFalcetaVishniac:2014}. Three different cuts
(one XY plane at Z=-0.1 and two YZ-planes at X=-0.25 and X=0.42) through the
computational domain show the strength of magnetic field $|\vec{B}|$ at the
evolution time $t=1.0$. Kelvin-Helmholtz-type structures are well seen in the
planes perpendicular to the reconnecting magnetic component $B_x$.  In the Z
direction, the Kelvin-Helmholtz instability is slightly suppressed by the guide
field of the strength $B_z=0.1 B_x$ (with $B_x=1.0$ initially). The initial
current sheet is located along the XZ plane at Y=0.0. A weak ($E_{kin} \approx
10^{-4} E_{mag}$) random velocity field was imposed initially in order to seed
the reconnection. \label{fig:bmag_cuts}}
\end{figure}

Recent simulations from \cite{KowalLazarianFalcetaVishniac:2014} are presented in
Figure~\ref{fig:bmag_cuts}.  The figure shows a few slices of the magnetic field
strength $|\vec{B}|$ through the three-dimensional computational domain with
dimensions $L_x=1.0$ and $L_y=L_z=0.25$.  The simulation was done with the
resolution $2048 \times 512 \times 512$.  Open boundary conditions along the X
and Y directions allowed studies of steady state turbulence.  At the presented
time $t=1.0$ the turbulence strength increased by two orders of magnitude from
its initial value of $E_{kin} \approx 10^{-4} E_{mag}$.  Initially, only the
seed velocity field at the smallest scales was imposed (a random velocity vector
was set for each cell).  We expect that most of the injected energy comes from
the Kelvin-Helmholtz instability induced by the local interactions between the
reconnection events, which dominates in the Z-direction, along which a weak
guide field is imposed ($B_z=0.1 B_x$).  As seen in the planes perpendicular to
$B_x$ in Figure~\ref{fig:bmag_cuts}, Kelvin-Helmholtz-like structures are
already well developed at time $t=1.0$. Turbulent structures are also observed
within the XY-plane, which probably are generated by the strong interactions of
the ejected plasma from the neighboring reconnection events.

\section{Mean field approach and turbulent reconnection}

The relation of turbulence and reconnection has attracted more attention
recently. For instance, \cite{GuoDiamondWang:2012} proposed a model based on the
earlier idea of mean field approach suggested initially in
\cite{KimDiamond:2001}.  In the latter paper the author concluded that the
reconnection rate should be always slow in the presence of turbulence. On the
contrary, models in \cite{GuoDiamondWang:2012} invoke hyperresistivity and get
fast reconnection rates. Similarly, invoking the mean field approach
\cite{HigashimoriHoshino:2012} presented their model of turbulent reconnection.

The mean field approach invoked in the aforementioned studies was critically
analyzed by \cite{EyinkLazarianVishniac:2011}, and below we briefly present some
arguments from that study. The principal difficulty is with the justification of
using the mean field approaches to explain fast magnetic reconnection. In such
an approach effects of turbulence are described using parameters such as
anisotropic turbulent magnetic diffusivity and hyper-resistivity experienced by
the fields once averaged over ensembles. The problem is that it is the lines of
the full magnetic field that must be rapidly reconnected, not just the lines of
the mean field.  \cite{EyinkLazarianVishniac:2011} stress that the former
implies the latter, but not conversely. No mean-field approach can claim to have
explained the observed rapid pace of magnetic reconnection unless it is shown
that the reconnection rates obtained in the theory are strictly independent of
the length and timescales of the averaging. More detailed discussion of the
conceptual problems of the hyper-resistivity concept and mean field approach to
magnetic reconnection is presented in \cite{LazarianVishniacCho:2004} and
\cite{EyinkLazarianVishniac:2011}.

\section{Summary}
The studies on turbulent magnetic reconnection has shown
\begin{itemize}
\item Turbulence makes reconnection fast, i.e. independent of resistivity.
\item Turbulence ubiquitous in astrophysics, but reconnection itself induces
turbulence.
\item Numerical simulations and observational data are consistent with the
turbulent reconnection being the dominant mechanism of astrophysical
reconnection.
\end{itemize}

%%==============================================================================
%%
\paragraph{Acknowledgements.}

A.L. research is supported by the NSF grant AST~1212096, Vilas Associate Award
as well as the support 1098 from the NSF Center for Magnetic Self-Organization.
The research is supported by the Center for Magnetic Self-Organization in
Laboratory and Astrophysical Plasmas. G.K. acknowledges support from FAPESP
(projects no. 2013/04073-2 and 2013/18815-0).

\bibliography{references}

\end{document}